\documentclass[aps,prd,amsmath,amssymb,twocolumn,preprintnumbers,nofootinbib]{revtex4}
\pdfoutput=1
\usepackage{graphicx}
\usepackage{color}
\usepackage{hyperref}					
\usepackage{mathtools}					
\usepackage{slashed}	
\usepackage{multirow}

\renewcommand{\epsilon}{\varepsilon}		
\newcommand{\beq}{\begin{eqnarray}}
\newcommand{\eeq}{\end{eqnarray}}

\newcommand{\bmp}{\noindent\begin{minipage}{16cm}}
\newcommand{\emp}{\end{minipage}\vskip 7mm} % 7mm untightened

\newcommand{\smc}[1]{ {\scriptstyle{\text{#1}}} }	%	Laver god cap tilsubscript	

\newcommand{\Nc}{N_\text{c}}
\newcommand{\Nf}{N_\text{f}}

\renewcommand{\vec}{\mathbf} 
 	
\DeclareMathOperator{\tr}{Tr}

\def\simge{\mathrel{%
   \rlap{\raise 0.511ex \hbox{$>$}}{\lower 0.511ex \hbox{$\sim$}}}}
\def\simle{\mathrel{
   \rlap{\raise 0.511ex \hbox{$<$}}{\lower 0.511ex \hbox{$\sim$}}}}

\usepackage{dcolumn}	% Align table columns on decimal point
\usepackage{bm}		% bold math
\usepackage{bbm}
\usepackage{subfigure}
\usepackage{pxfonts}

\definecolor{rossoCP3}{cmyk}{0,.88,.77,.40}
\definecolor{light_gray}{rgb}{0.8,0.8,0.8}

\def\lsim{\mathrel{\rlap{\lower4pt\hbox{\hskip1pt$\sim$}}
    \raise1pt\hbox{$<$}}}                % less than or approx. symbol
\def\gsim{\mathrel{\rlap{\lower4pt\hbox{\hskip1pt$\sim$}}
    \raise1pt\hbox{$>$}}}                % greater than or approx. symbol

\newcommand{\be}{\begin{eqnarray}}
\newcommand{\ee}{\end{eqnarray}}

\newcommand{\Tr}{\textrm{Tr}}

\newcommand{\eariv}[1]{\left\langle #1 \right\rangle}

\begin{document}
%%%%%%%%%%%%%%%%%%%%%%%%%%%%%%%%%%%%%%%%%%%%%%%%%%%%%%%%%%%%%%%%%%%%%%%%%%%
%\includegraphics[width=3.cm]{CP3-logo} 
\title{\Large  \color{rossoCP3}  Orthogonal Technicolor with Isotriplet Dark Matter  on the Lattice}
\author{Ari {\sc Hietanen}$^{\color{rossoCP3}{\varheartsuit}}$}
\email{hietanen@cp3-origins.net} 
\author{Claudio {\sc Pica}$^{\color{rossoCP3}{\varheartsuit}}$}
\email{pica@cp3-origins.net}
\author{Francesco {\sc Sannino}$^{\color{rossoCP3}{\varheartsuit}}$}
\email{sannino@cp3.dias.sdu.dk} 
\author{Ulrik Ish\o j {\sc S\o ndergaard}$^{\color{rossoCP3}{\varheartsuit}}$}
\email{sondergaard@cp3.sdu.dk} 
\affiliation{
$^{\color{rossoCP3}{\varheartsuit}}$CP$^3$-Origins \& the Danish Institute for Advanced Study DIAS,
        University of Southern Denmark, Campusvej 55, DK-5230 Odense M, Denmark.}
%\preprint{CP3-Origins-2012-030 \& DIAS-2012-31}
%%%%%%%%%%%%%%%%%%%%%%%%%%%%%%%%%%%%%%%%%%%%%%%%%%%%%%%%%%%%%%%%%%%%%%%%%%%%%%%%%%%%%%%%
\begin{abstract}
We study the gauge dynamics of an SO(4)-gauge theory with two Dirac Wilson fermions transforming according to the vector representation of the gauge group. We determine the lattice phase diagram by locating the strong coupling bulk phase transition line and the zero PCAC mass line. We present results for the spectrum of the theory. In particular we measure the pseudoscalar, vector and axial meson masses. The data are consistent with a chiral symmetry breaking scenario rather than a conformal one. 
When used to break the electroweak symmetry dynamically the model leads to a natural dark matter candidate.
% phenomenological relevance for the theory relies on the fact that the pattern of chiral symmetry breaking is identical to the one of Minimal Walking Technicolor model albeit with relevant distinctive features: It provides a complex weak isotriplet of Goldstone bosons of which the neutral component can be identified with a light composite dark matter state; b] It is expected to break the global symmetry spontaneously; c] It is free from fermionic composite states made by a techniglue and a technifermion.\
\\[.1cm]
{\footnotesize  \it Preprint: CP3-Origins-2012-030 \& DIAS-2012-31}
\end{abstract}

\maketitle

%%%%%%%%%%%%%%%%%%%%%%%%%%%%%%%%%%%%%%%%%%%%%%%%%%%%%%%%%
\section{Introduction}
{ Understanding the phase diagram of strongly interacting theories will unveil a large number of theories of fundamental interactions useful to describe electroweak symmetry breaking, dark matter and even inflation \cite{Sannino:2008ha,Channuie:2011rq,Bezrukov:2011mv,Channuie:2012bv}.
To gain a coherent understanding of strong dynamics  besides the SU($N$) gauge groups \cite{Sannino:2004qp,Dietrich:2006cm}, one should also investigate the orthogonal, symplectic and exceptional groups.
SO($N$) and SP($2N$) phase diagrams were investigated with analytic methods in \cite{Sannino:2009aw}, while the exceptional ones together with orthogonal gauge groups featuring spinorial matter representations were studied in \cite{Mojaza:2012zd}. So far lattice simulations have been mostly employed to explore the phase diagram of SU($N$) gauge theories while a systematic lattice analysis of the smallest symplectic group  was launched in \cite{Lewis:2011zb}. 

Here we move forward by analyzing on the lattice the dynamics of the SO($4$) gauge group with two Dirac fermions in the vector representation of the group.  This choice is based on the following theoretical and phenomenological considerations. The theory is expected to be below or near the lower boundary of the conformal window \cite{Sannino:2009aw,Frandsen:2009mi}, and therefore break chiral symmetry. The theory can be used as a technicolor \cite{Weinberg:1979bn,Susskind:1978ms} template similar, from the global symmetry point of view, to Minimal Walking Technicolor (MWT) \cite{Sannino:2004qp,Dietrich:2005jn,Foadi:2007ue}. However, the fermion representation is such that, differently from MWT, one cannot construct composite fermions out of one techniquark and one techniglue. This removes immediately the presence of fractionally charged states with the simplest choice of the hypercharge assignment. Furthermore the technicolor theory leads to a weak isotriplet with the neutral member being an ideal dark matter candidate \cite{Sannino:2009aw,Frandsen:2009mi}, the isotriplet Technicolor Interactive Massive Particle (iTIMP). This state is a pseudo Goldstone and therefore can be light with respect of the electroweak scale making it an natural candidate to resolve some of the current experimental puzzles \cite{Frandsen:2009mi,DelNobile:2011je}.  The first model featuring composite dark matter pions appeared in \cite{Gudnason:2006ug,Ryttov:2008xe} and the first study of technipion dark matter on a lattice appeared in \cite{Lewis:2011zb}.

Due to the reality of the fermion representation the quantum global symmetry group is SU(4) expected to break spontaneously to SO(4), yielding nine Goldstone bosons. Once gauged under the electroweak theory three are eaten by the SM gauge bosons. Six additional Goldstone bosons form an electroweak complex triplet of technibaryon with the neutral isospin zero component to be identified with the iTIMP of \cite{Frandsen:2009mi}.

SO(4) is a semi simple group, SO(4) $\cong$ SU(2)$\otimes$SO(3), and it has a non-trivial center $Z_2$.  The theory is asymptotically free and since the two-loop $\beta$-function for different number of flavors looses the infrared zero for $N_f = 2.3$ while the all-orders beta function \cite{Ryttov:2007cx,Pica:2010mt} predicts the anomalous dimension of the mass to be unity for $N_f =2.86$ we expect that chiral symmetry breaks for two Dirac flavors. However,  we want to confirm here this result via first principle lattice simulations. Furthermore  there is also the possibility that the theory shows a certain degree of walking \cite{Sannino:2009aw,Frandsen:2009mi,Holdom:1981rm,Yamawaki:1985zg,Appelquist:1986an} unless the phase transition is of jumping type \cite{Sannino:2012wy,deForcrand:2012vh}. Jumping conformal phase transitions have been demonstrated to occur in a wide class of theories \cite{Antipin:2012sm}.

As a natural first step, we study the phase diagram in the ($\beta,m_0$)-plane to find the relevant region of parameter space to simulate. We then determin the zero PCAC mass line as well as the strong coupling bulk phase transition line. In addition, we report on the pseudoscalar, vector and axial vector meson masses. From the measured spectrum we infer that the  theory breaks chiral symmetry dynamically. Part of these results appeared in \cite{Hietanen:2012qd}.

In Section~\ref{theory} we present the analytic expectations for the phase diagram of SO($N$) as function of the number of flavors. We also summarize the expected breaking pattern of the quantum global symmetries for theories below the conformal window.  We also prove the spectral degeneracy between certain diquarks and ordinary meson-like states. In Section~\ref{lattice} we recall the lattice formulation of the theory and summarize the physical observable studied here. The results of the simulations are reported in Sect.~\ref{results} and conclude in Sect.~\ref{conclusion}.

\section{Orthogonal conformal window and Chiral Symmetry breaking pattern} \label{theory}
 The two loop $\beta$-function for an SO($N$) theory with $\Nf$ Dirac fermions transforming according to the vector representation of the gauge group is
\begin{equation}
  \beta(\alpha)=-\frac{\alpha^2}{2\pi}\left(  b_0 + b_1\frac{\alpha}{2\pi} \right) \, ,
\end{equation}
where
\begin{align}
\begin{split}
b_0 &= \frac{11}{3}\Nc - \frac{4}{3}\Nf - \frac{22}{3} \, , \\
b_1&=-\frac{10}{3} (\Nc-2) \Nf-(\Nc-1) \Nf+\frac{17}{3}
   (\Nc-2)^2 \, .
\end{split}
\end{align}
A naive estimate of the lower bound of conformal window is given when the second coefficient $b_1$ changes sign. For SO(4) this happens when $N_f = \frac{68}{29}\simeq 2.3$. The corresponding values for three and four-loops in the $\overline{\text{MS}}$-scheme are $N_f=1.8$ and $N_f=3.0$. The all-orders beta function predicts as lower boundary $N_f=2.86$, see Fig.~\ref{conformalwindow}. Hence, perturbative and nonperturbative methods suggest that chiral symmetry breaks for two Dirac flavors. However, lattice simulations can seal this expectation. 
\begin{figure}[tb]
  \begin{center}
  \vspace{10pt}
  \includegraphics[width=0.9\columnwidth]{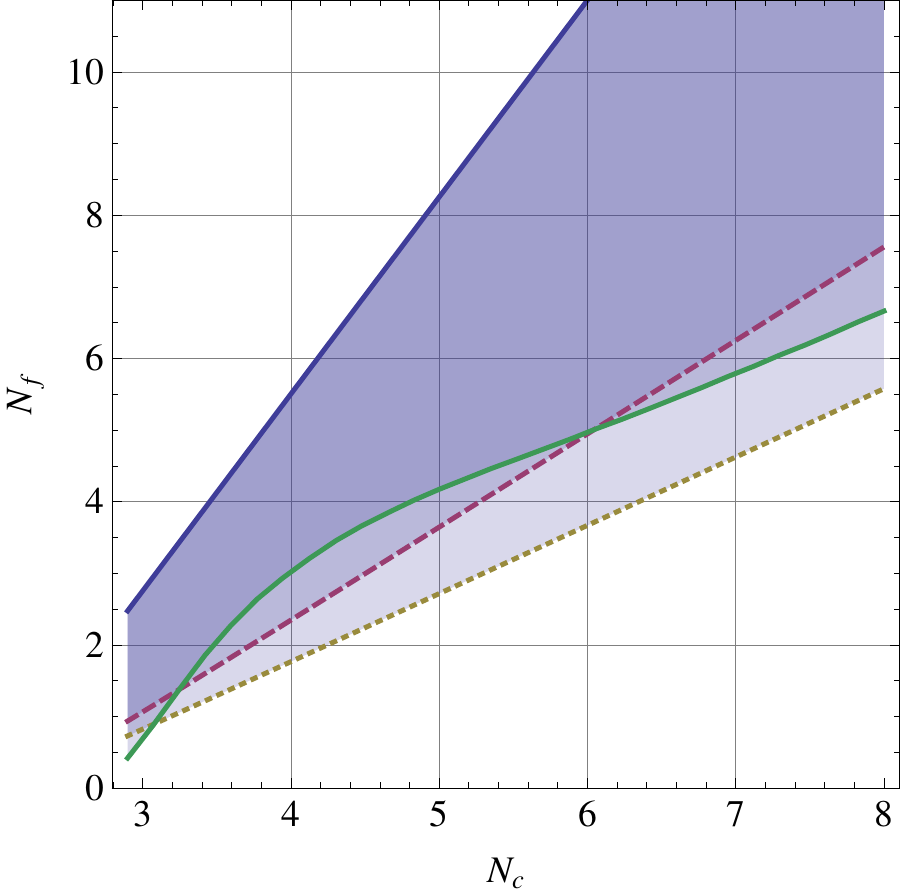}
    \caption{
    Conformal window of SO($\Nc$) with $\Nf$ Dirac fermions in the fundamental representation. Upper bound is when asymptotic freedom is lost. Lower bounds are 2-loop (red, dashed), 3-loop (yellow, dotted) and 4-loop estimates (green, solid).
    \label{conformalwindow}} 
  \end{center}
\end{figure}
Since the vector representations of orthogonal groups are real the quantum global symmetry of the theory is, for a generic $N_f$  SU(2$\Nf$) which is larger than SU($\Nf$)$\otimes$SU($\Nf$)$\otimes$ U$_V(1)$  valid for complex  fermion representations.  
The reality property of the representation translates in the following property of the Dirac operator
\begin{equation}
(\slashed D+m) C\gamma^5 = C \gamma^5 (\slashed D + m)^*, \label{DiracMatrixSymmetry}
\end{equation}
where $\slashed D=\gamma^\mu (\partial_\mu-i g A^a_\mu \tau_a)$, $a=1,..., d[G]$ where $d[G]$ is the dimension of the adjoint representation of the gauge group and $C=i \gamma^0 \gamma^2$ is the charge conjugation operator. 

The global SU(2$\Nf$)  is assumed to break to the maximal diagonal subgroup 
\begin{equation}
\text{SU}(2\Nf) \quad \rightarrow \quad \text{SO}(2\Nf ) \, ,
\end{equation}
for the massless theory and for $N_f$ below the conformal window. A common mass for the Dirac fermions leads to the same pattern of explicit symmetry breaking. The explicit interpolating operators for the Goldstones can be naturally divided in three independent antifermion-fermion bilinears 
\begin{equation}
\bar \psi_{f} \gamma^5 \psi_{f^\prime}\, ,
\end{equation}
with $f$ and $f^{\prime}$ the flavor indices $f = 1,2$ and six {\it difermion} operators  
\begin{equation}
\psi_f ^TC\gamma^5 \psi_{f^{\prime}} \qquad \text{and}\qquad \bar \psi_{f} \gamma^5 C \bar \psi^{T}_{f^{\prime}} \, .
\end{equation}
 The reader can find a useful summary of the global symmetry breaking patterns tailored for lattice computations in \cite{Kogut:2000ek} while applications to beyond standard model physics for similar patterns appeared in \cite{Appelquist:1999dq,Gudnason:2006ug}. 
Notice that whereas the usual pions have odd parity, the corresponding diquarks are parity even. 
%We will argue, that in the presence of a bare mass term the pseudo Nambu-Goldstone bosons will remain degenerate. 
It was noticed in \cite{Lewis:2011zb} that when fermions are in a pseudoreal representation, the diquark correlator is exactly identical to the corresponding mesonic correlator. In appendix \ref{AppCorrelator} we give a similar proof for fermions in real representations. The proof uses the symmetry \eqref{DiracMatrixSymmetry} of the Dirac operator along with the $\gamma^5$-hermiticity $\gamma^5(\slashed D+m) \gamma^5 = (\slashed D + m)^\dagger$ property. The result can be stated as
\begin{equation}
c_{\bar \psi_f \bar \psi_{f^{\prime}}}^{(\Gamma )}(x-y)=c_{\bar \psi_f \psi_{f^\prime}}^{(\Gamma )}(x-y) = c_{\psi_f \psi_{f^\prime}}^{(\Gamma )}(x-y) \label{CorrId}
\, ,
\end{equation}
where $c_{\bar \psi_f \psi_{f^\prime}}^{(\Gamma )}$ is the correlator for the operator $\bar \psi_f \Gamma \psi_{f^{\prime}}$ and $c_{\psi_f \psi_{f^{\prime}}}^{(\Gamma )}$ is the correlator for the corresponding diquark operator $\psi_f^T \Gamma C \psi_{f^{\prime}}$. $\Gamma$ can be any of the matrices $\mathbf{1}, \gamma^5 , \gamma^\mu, \gamma^\mu \gamma^5$. 

Having discussed the generic features expected for orthogonal groups we now turn to the lattice formulation and results for the relevant case of  SO($4$) with two Dirac flavors.

 \section{ Lattice formulation}
 \label{lattice}
In this work we have used the Wilson prescription for the lattice action
\begin{equation}
S=S_\smc{F} +S_\smc{G}\, ,
\end{equation}
where
\begin{equation}
S_\smc{G}= \beta \sum_x \sum_{\mu,\nu < \mu} \left[  1- \frac{1}{\Nc} \tr U_{\mu\nu}(x) \right] \, , \label{eq:Sg}
\end{equation}
is the Yang-Mills gauge action.  We have normalized the lattice spacing to $a=1$. $U_{\mu\nu}(x)$ is the plaquette defined in terms of the link variables as
\begin{equation}
U_{\mu\nu}(x)=U_\mu(x)U_\nu(x+\hat \mu)U^T_\mu(x+\hat \mu+\hat \nu)U^T_\nu(x+\hat \nu) \, .
\end{equation}
The Wilson fermion action is \begin{equation}
S_\smc{F}=\sum_{f} \sum_{x,y}  \bar \psi_f(x) M(x,y) \psi_f(y) \ ,
\end{equation}
with $f$ running over fermion flavors and the Wilson-Dirac matrix $M(x,y)$ given by
\begin{align}
\begin{split}
\sum_{y}M&(x,y) \psi(y) = (4+m_0)\psi (x)\\
& - \frac{1}{2}\sum_{\mu}\Big[(1+\gamma_\mu)U^T_\mu(x-\hat\mu)\psi(x-\hat\mu)  \\
&\qquad\qquad\quad+   (1-\gamma_\mu)U_\mu(x)\psi(x+\hat\mu) \Big] \, .
\end{split} \label{eq:Wilson-Dirac}
\end{align}
Here the gauge and spinor indices have been suppressed. The bare parameters are the inverse of the bare coupling $\beta=2\Nc/g_0^2$ appearing in the gauge action  and the bare mass $m_0$ of the Wilson fermions. 

We employ the Partial Conservation of the Axial Current (PCAC) relation to define the physical quark mass
\begin{equation}
  m_{\rm PCAC}=\lim_{t \rightarrow \infty}\frac{1}{2}\frac{\partial_t V_{\rm PS}}{V_{\rm PP}},
\label{eq:PCAC}
\end{equation}
where the currents are
\begin{align}
  V_{\rm PS}(x_0) &= a^3\sum_{x_1,x_2,x_3} \eariv{\bar{\psi}_1(x)\gamma_0 \gamma_5 \psi_2(x)\bar{\psi}_1(0)\gamma_5\psi_2(0)} \ ,\nonumber \\
  V_{\rm PP}(x_0) &= a^3\sum_{x_1,x_2,x_3} \eariv{\bar{\psi}_1(x)\gamma_5\psi_2(x)\bar{\psi}_1(0)\gamma_5\psi_2(0)} \ .
\end{align}

The meson masses are estimated using time slice averaged zero momentum correlators 
\begin{equation}
C_{\bar \psi_1\psi_2}^{(\Gamma )}(x_0)= a^3 \sum_{x_1,x_2,x_3}\Tr \left( \left[  \bar \psi_1(x) \Gamma \psi_2(x) \right]^\dagger  \bar \psi_1(0) \Gamma \psi_2(0)  \right)\ ,
\end{equation}
where $\Gamma=\gamma_5$ for pseudoscalar, $\Gamma=\gamma_k$ ($k=1,2,3$) for vector, and $\gamma_5\gamma_k$ for axial vector meson.

\section{Results}
\label{results}

%%%%%%%%%%%%%%%% FIGURE: THERMALIZATION
%\begin{figure}[tbp]
%\begin{center}
%\includegraphics[width=1\columnwidth]{ThermOfLowEig.pdf}
%\caption{Thermalization of the lowest eigenvalue $\lambda_0$ of squared Dirac operator as a function of iterations on a $24^3\times64$ lattice with $\beta=7$.}
%\label{fig:thermalization}
%\end{center}
%\end{figure}
%%

%%%%%%%%%%%%%%%% FIGURE: LATTICE PHASES
\begin{figure}[tbp]
\begin{center}
\includegraphics[width=1\columnwidth]{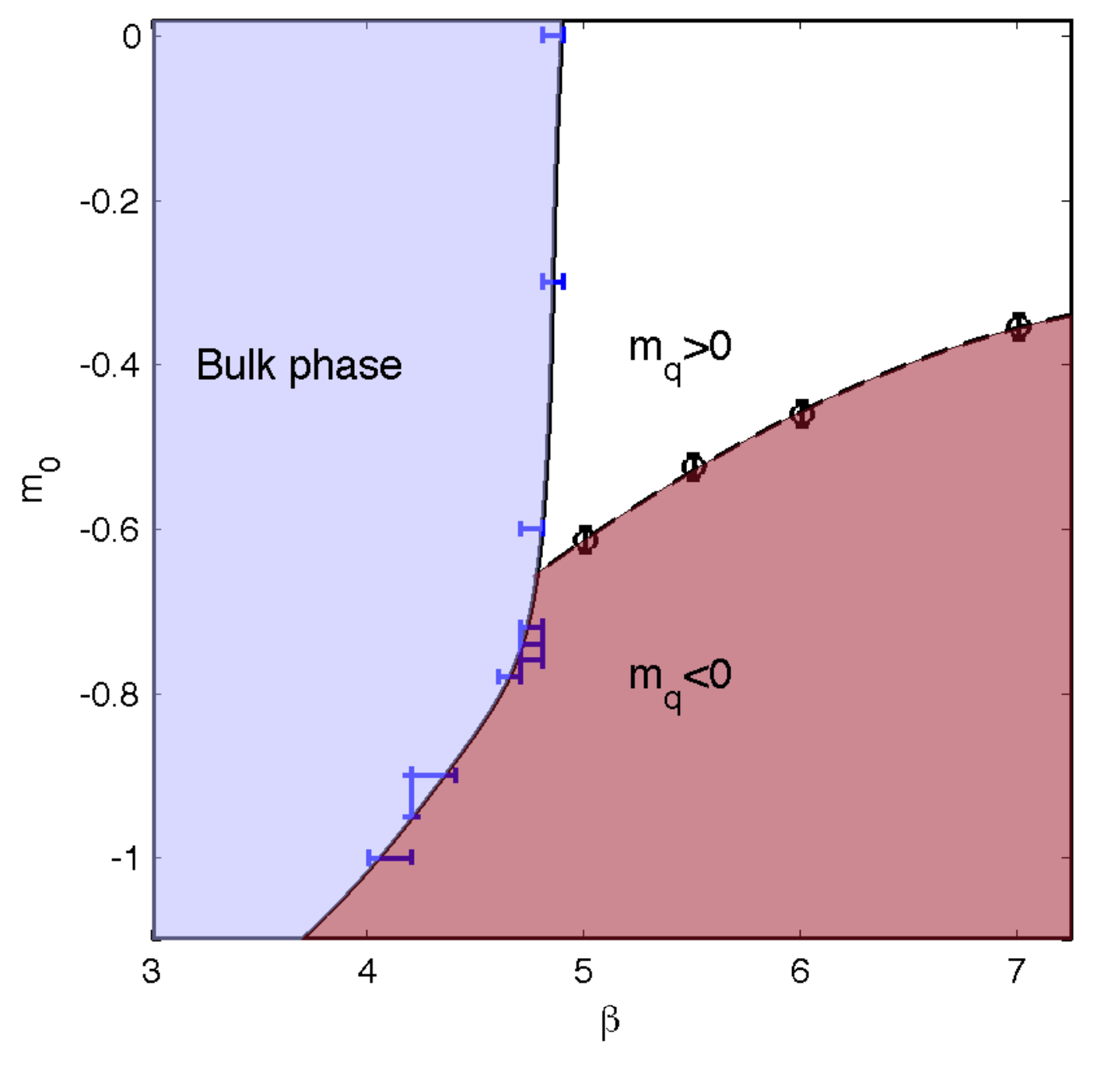}
\caption{Lattice phase structure outlined on an $8^3\times16$ lattice. Circles represent points of critical bare mass where $m_\smc{PCAC}=0$. The transition between the bulk phase is of first order. The error bars represent the interval over which the measured average plaquette jumps. }
\label{fig:lattice_phases}
\end{center}
\end{figure}
%
%%%%%%%%%%%%%%%% FIGURE: PLAQUETTE VS beta
%% \begin{figure*}[bt]
%% \begin{minipage}[b]{0.45\linewidth}
%% \centering
%% \includegraphics[width=\textwidth]{PlaquetteVSbeta.pdf}
%% \caption{Average plaquette $\langle P \rangle$ vs. $\beta$ on an $8^3\times16$ lattice at three different values of the bare mass.}
%% \label{fig:plaqVSbeta}
%% \end{minipage}
%% \hspace{0.5cm}
%% \begin{minipage}[b]{0.45\linewidth}
%% \centering
%% \includegraphics[width=\textwidth]{m0_mpcac_b7.pdf}
%% \caption{Physical quark mass measured from the PCAC relation vs. bare mass $m_0$ at fixed $\beta=7$.}
%% \label{fig:pcacVSm0}
%% \end{minipage}
%% \end{figure*}

\begin{figure}[bt]
\begin{center}
\includegraphics[width=\columnwidth]{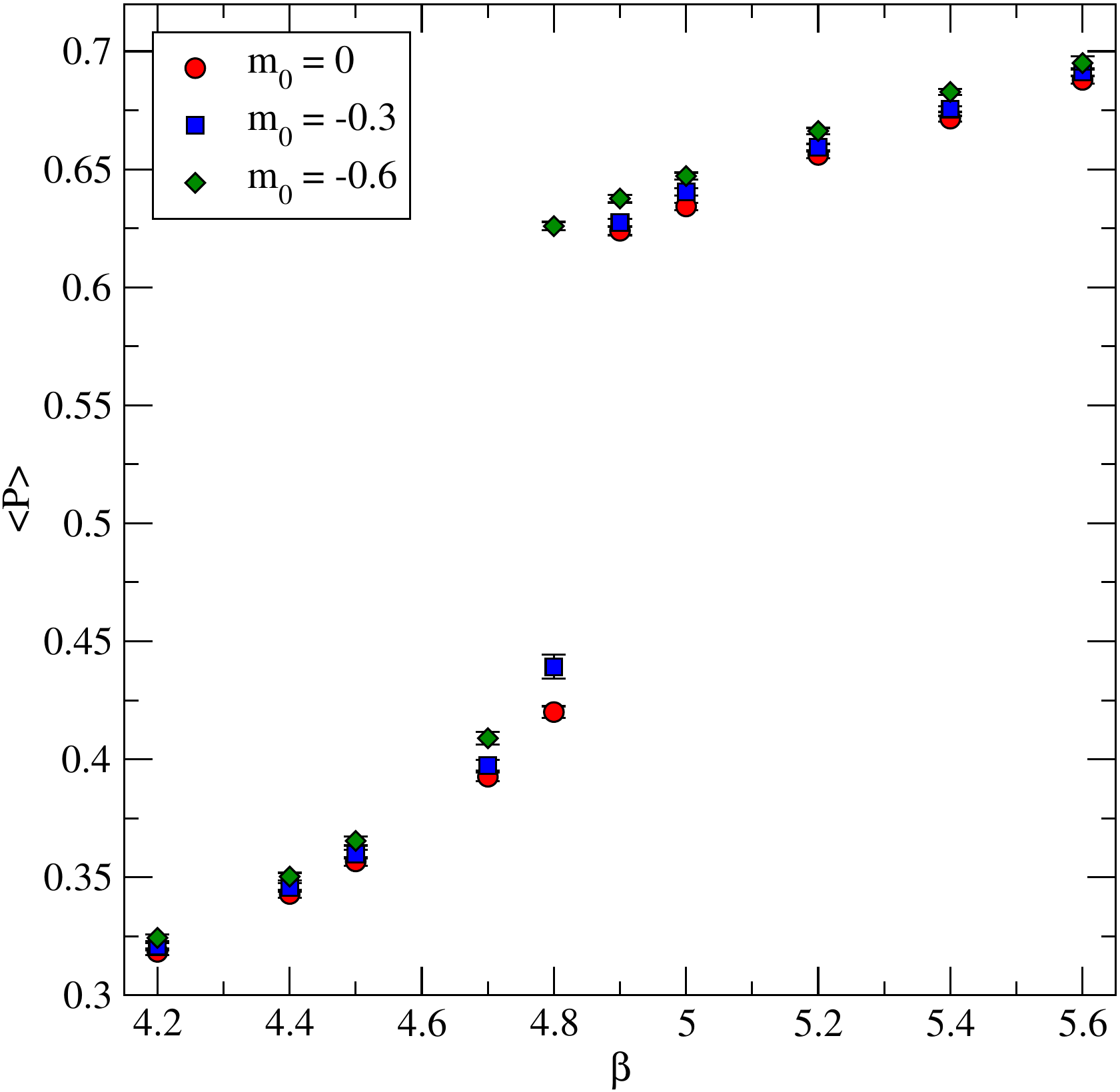}
\caption{Average plaquette $\langle P \rangle$ vs. $\beta$ on an $8^3\times16$ lattice at three different values of the bare mass. \label{fig:plaqVSbeta}}
\end{center}
\end{figure}

%\begin{figure}
%\begin{center}
%\includegraphics[width=\columnwidth]{m0_mpcac_b7.pdf}
%\caption{Physical quark mass measured from the PCAC relation vs. bare mass $m_0$ at fixed $\beta=7$. \label{fig:pcacVSm0}}
%\end{center}
%\end{figure}
%%

\begin{table}
  \begin{tabular}{cccc}
    \hline
    Volume & $\beta$ & Iterations & Thermalization \\
    \hline
    \hline
    \multirow{2}{*}{$8^3\times16$} & 4.1,4.2\dots 4.9, 5.2, 5.4, 5.6 & 2000 & 500 \\
     & 4 4.5 5, 5.5, 6, 7 & 5000 & 2000 \\
    $12^3\times64$ & 5.5, 7 & 5000 & 1500\\
    $24^3\times64$ & 7 & 850 - 2000 & 600 \\
    \hline
  \end{tabular}
  \caption{Simulation parameters and thermalization times. For each coupling we performed multiple simulations with appropriate bare masses. The thermalization  column refers to the number of discarded initial configurations. \label{table:sim_param}}
\end{table}

The simulations were performed on three different lattices $8^3 \times 16$, $12^3 \times 64$ and $24^3 \times 64$ where in all cases the larger dimension is the temporal one. All the simulations were started from a random configuration and the first 500-2000 iterations were discarded. This is enough to thermalize the system for the quantities we measured. 
%the lowest eigenvalue of the square Dirac %operator (see Fig.~\ref{fig:thermalization}), %maybe except for  $m_0=-0.4$ for a volume %of $24^3 \times 64$, which is extremely %expensive to simulate. 
For a complete list of the simulations see Table~\ref{table:sim_param} where we have omitted the values of the bare masses.
 
The smallest lattice was used for exploration of the parameter space spanned by the bare mass $m_0$ and the coupling $\beta$. Fig.~\ref{fig:lattice_phases} shows an outline of the lattice phase structure measured on this $8^3\times 16$ lattice. For small values of $\beta$ the system is in a bulk phase not connected to continuum physics. The bulk phase is separated from the small coupling (large $\beta$) phase by a first order phase transition. Fig.~\ref{fig:plaqVSbeta} shows the discontinuous behavior of the average plaquette when crossing the bulk phase transition, for three different values of $m_0$. The uncertainty on the location of the bulk phase transition shown in Fig.~\ref{fig:lattice_phases} is due to taking discrete values of $\beta$ between simulation points.

We can compare our result for the location of the bulk transition to previous studies of SO(N) pure gauge theories. Earlier simualtions focused mainly on the SO(3) gauge group~\cite{deForcrand:2002vs} with the exception of~\cite{Bursa:2012ab} where also other values of $N$ were considered.
For SO(4) the authors of~\cite{Bursa:2012ab} find that the bulk phase transition happens for $4.62(3) < \beta < 4.87(3)$, which is in agreement with our result in Fig.~\ref{fig:plaqVSbeta}.
%The ranges of $\beta$ over which the average plaquette jumps are represented by error bars on the bulk phase transition line in Fig.~\ref{fig:lattice_phases}.

The critical line where the physical quark mass vanishes is determined from the PCAC relation \eqref{eq:PCAC}. The critical line of $m_q=0$ in the phase diagram (Fig.~\ref{fig:lattice_phases}) is constructed by linear fits to the PCAC mass. Fig.~\ref{fig:L8T16mpcac} shows the bare mass dependency of the PCAC mass at three different couplings on the $8^3\times 16$ lattice. 
%%%%%%%%%%%%%%%% MPCAC FIG
\begin{figure}[bt]
\begin{center}
\includegraphics[width=\columnwidth]{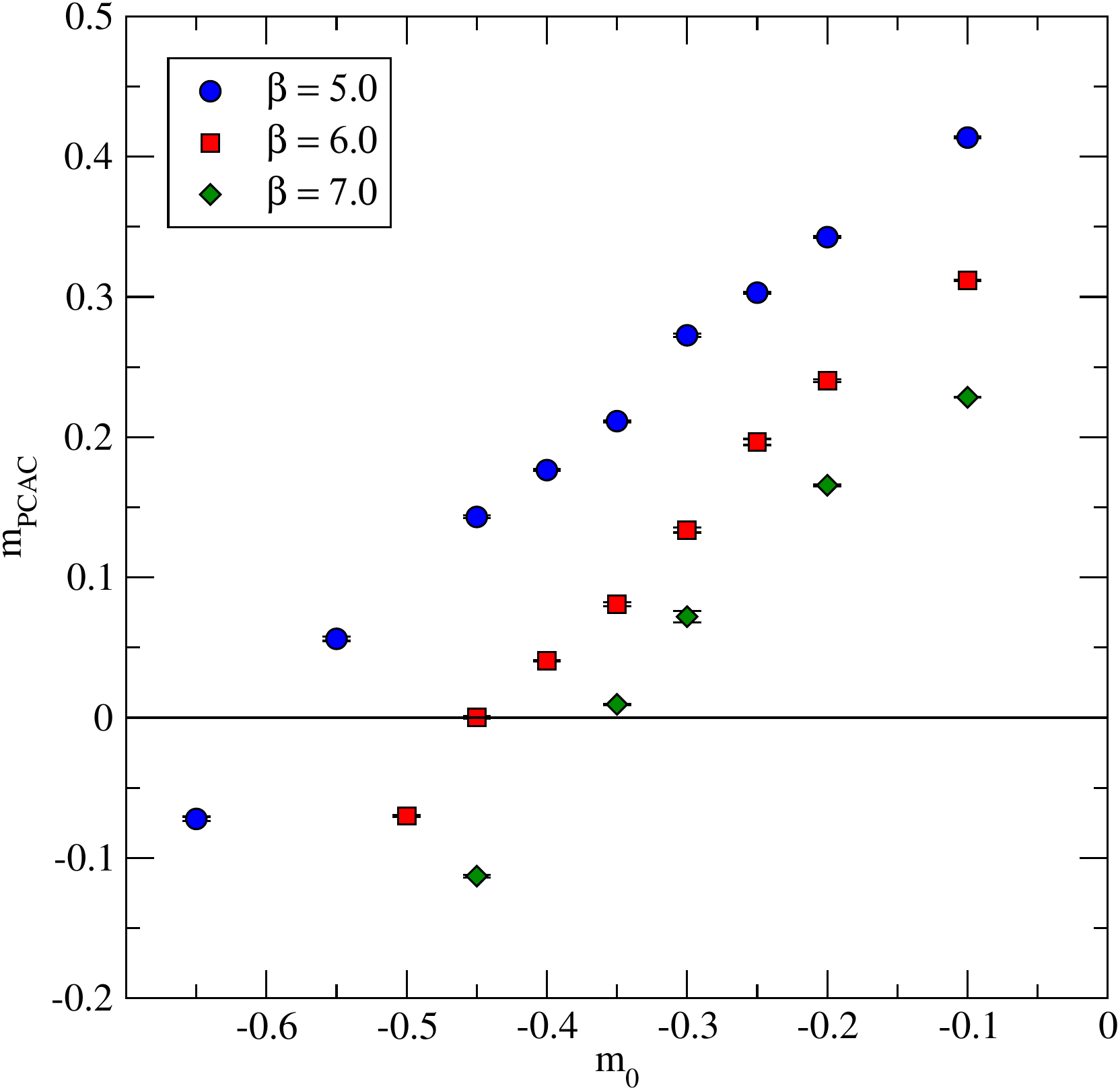}
\caption{$m_\smc{PCAC}$ in units of inverse lattice spacing at three different couplings. The measurements are performed on a $8^3\times 16$.}
\label{fig:L8T16mpcac}
\end{center}
\end{figure}
%%%%%%%
%%%%%%%%%%%%%%%% FINITE VOLUME FIGURE
\begin{figure}[bt]
\begin{center}
\includegraphics[width=\columnwidth]{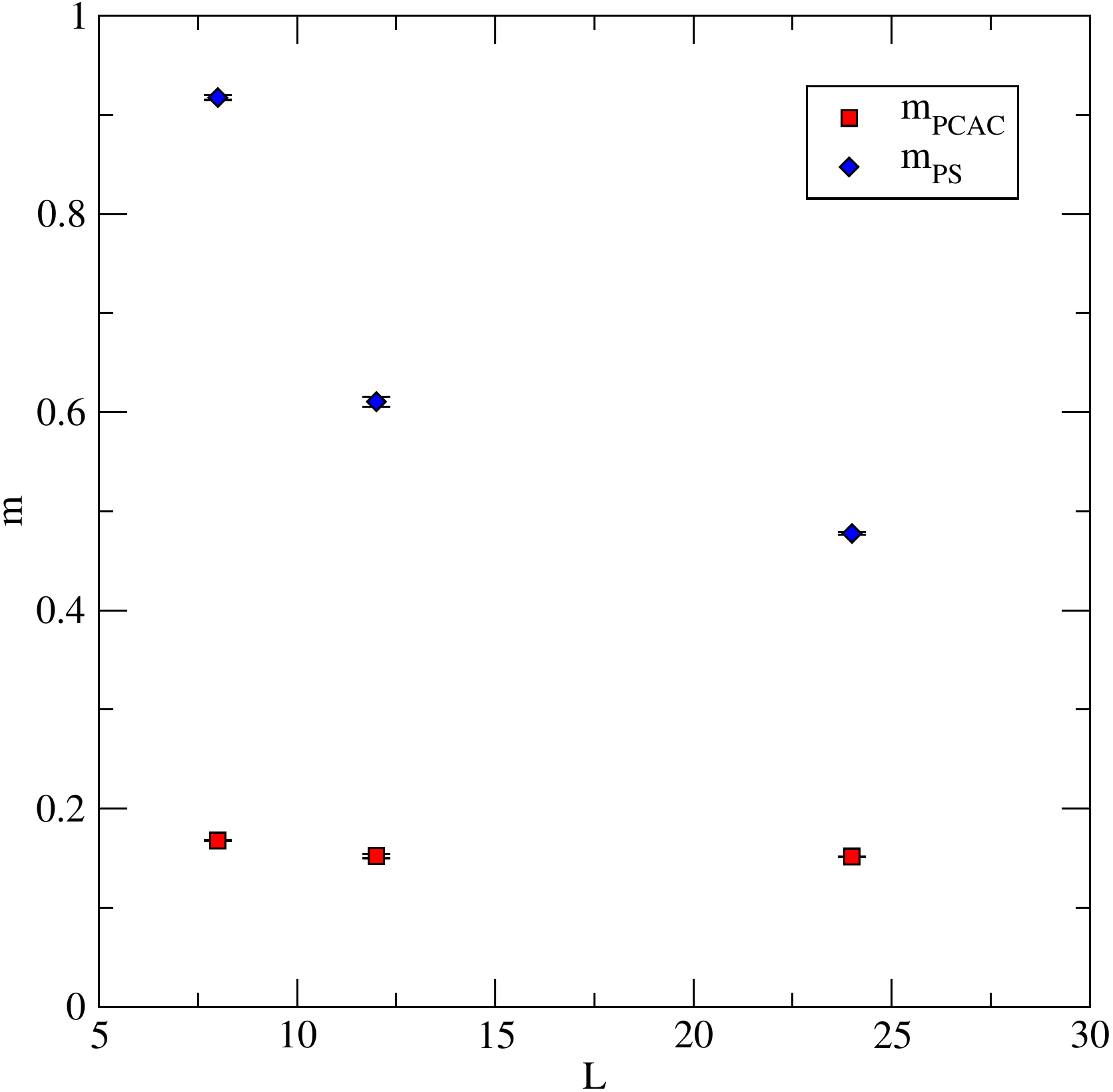}
\caption{Finite size effects on $m_\smc{PS}$ and $m_\smc{PCAC}$ in units of inverse lattice spacing. The measurements are performed on a $24^3\times 64$ lattice at $\beta=7$ and $m_0=-0.2$.}
\label{fig:finitevolume}
\end{center}
\end{figure}
%%%%%%%
\begin{figure}[bt]
\begin{center}
\includegraphics[width=\columnwidth]{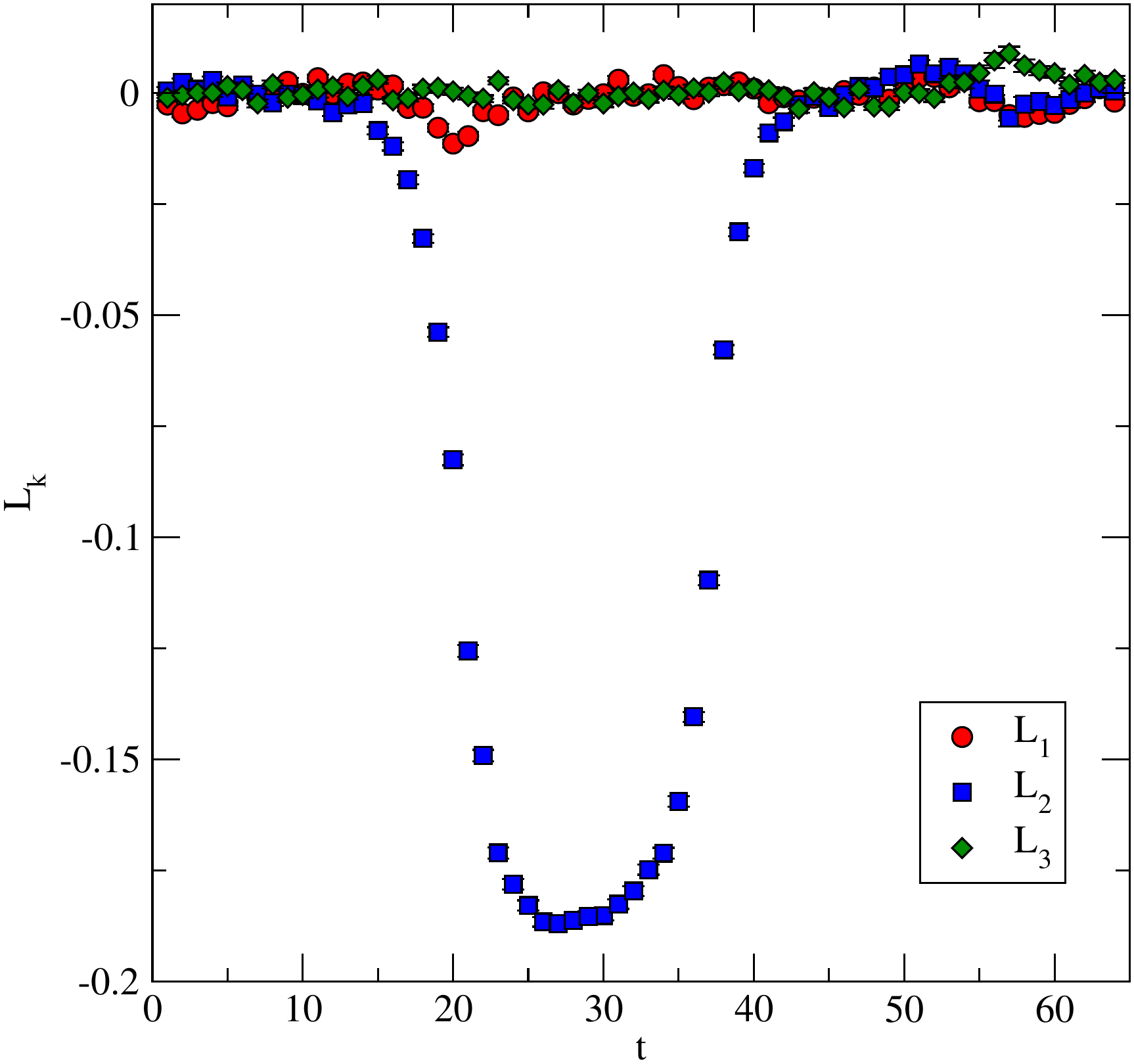}
\caption{Average Polyakov loops wound around the three spatial dimensions computed at each timeslice of the lattice. This measurement was performed on a $12^3\times64$ lattice at $\beta=7$ and $m_0=-0.3$. The values are averages over 700 configurations starting at 1800 where the system does not appear to thermalize further.}
\label{fig:poly_in_time}
\end{center}
\end{figure}
%%%%%%%%%%%%%%%%%%%%%%%%%%%%%%%%%%%
%
\begin{figure}[bt]
\begin{center}
\includegraphics[width=\columnwidth]{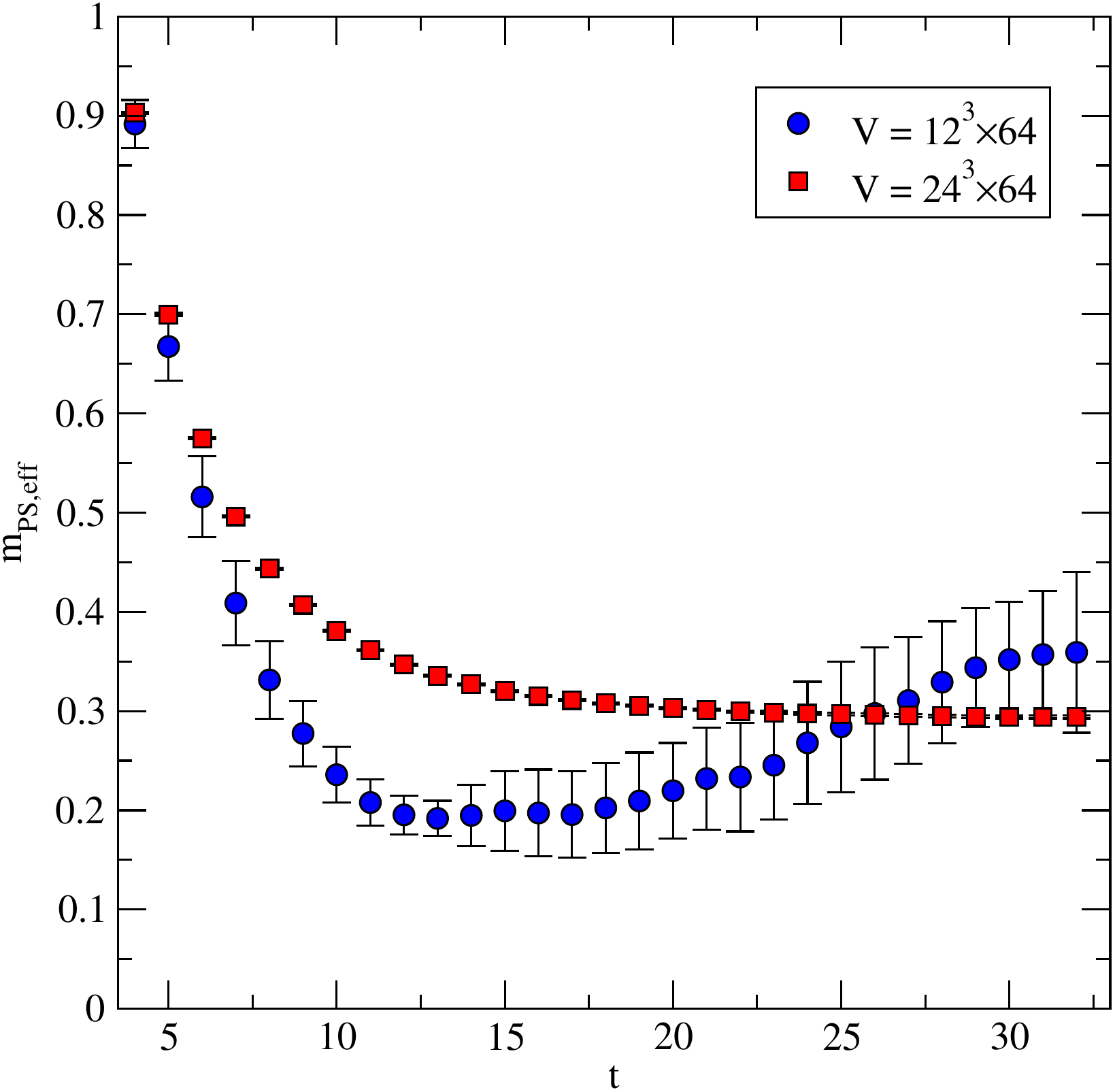}
\caption{Effective mass of pseudoscalar meson for two different volumes.}
\label{fig:eff_mass}
\end{center}
\end{figure}
%%%%%%%%%%%%%%%%%%%%%%%%%%%%%%%%%%%

\subsection{Finite size effects}
According to the perturbative estimates discussed in section \ref{theory} the running of the gauge coupling is expected to be slow. This also suggests that attention should be paid to finite size effects, which need to be estimated non-perturbatively by measuring physical observables as a function of lattice size. 

In the case of SO(N) pure gauge theories~\cite{deForcrand:2002vs,Bursa:2012ab} the bulk phase transition occurs at such a weak coupling that extremely large lattices are required for simulations in the confined phase, the one connected to the continuum physics.
However, in the presence of dynamical quarks, we find that somewhat smaller volumes ($24^3\times64$) are enough to probe the chiral regime of the system.

In figure Fig.~\ref{fig:finitevolume} the mass of the pseudoscalar meson $m_\smc{PS}$ and the PCAC quark mass $m_\smc{PCAC}$ is plotted for different lattice sizes. The PCAC mass has little dependence on the lattice size being a UV quantity.
The pseudoscalar meson mass, on the contrary, is very sensitive to finite size effects even if it is still somewhat heavy at the bare mass used in Fig.~\ref{fig:finitevolume}. 
%The observed $m_\smc{PS}$ seems to have stabilized when reaching $L \gtrsim 20$. We therefore proceed in the investigation of chiral symmetry using only the $24^3\times64$ lattice. 

Another interesting property which occurs in this model is a novel phase separation at small volumes. 

We observe the coexistence of two distinct phases which can be characterized by the average value of Polyakov loops wrapping around the three spatial directions taken on each time slice separately. In detail, the operators we consider are defined as
\begin{equation}
  L_k(t) = \left\langle\frac{1}{N_{i}N_{j}}\sum_{x_i,x_j} \frac{1}{N_{\textrm c}} \tr \prod_{x_k} U_k(t,\vec{x})\right\rangle,
\end{equation}
where $i\ne j \ne k$ are spatial directions.
 Fig.~\ref{fig:poly_in_time} shows the time resolved Polyakov loops on a $12^3\times 64$ lattice at $\beta=7$ and $m_0=-0.3$. The averaging is performed over 700 configurations belonging to the same simulation. The coexistence of two phases with different values of $L_2$ is clear from the figure. The phenomenon appears in all simulations performed on small lattices. The location of the phase boundaries and the direction in which the Polyakov loop has non-zero average is random. In some cases more than two phase boundaries appear in the same system. 

The coexistence of two phases with different mesonic correlation lengths separated by a domain walls could explain the unusual decorrelation of mesonic operators observed for volumes smaller than  $24^3 \times 64$. This is reflected in the rise of the effective mass plateaux of the pseudoscalar meson shown in Fig.~\ref{fig:eff_mass}.

In order to understand whether these phase separations are related to the presence of dynamical fermions we have also performed pure gauge simulations on $12^3\times 64$ lattices. The phase separation occurs also for the pure gauge. Thus the phenomenon seems to be a feature stemming from the pure gauge sector.

We will not explore this feature further, but it would be interesting to continue its investigation in the future.

% {\color{red} It will take some writing to state all the topological differences (N even, N odd, $N=4$, N multiple of 4 and so on)}

In order to avoid the complications stemming from the phase separation described above we use $24^3\times64$ lattices for the rest of the paper.

%%%%%%%%%%%%%%%%%%%%%%%%%%%%%%%%%%%
\subsection{Spectrum and chiral symmetry breaking}
We address the dynamical fate of the chiral symmetries of the theory by determining the pseudoscalars and (axial) vectors spectrum. 

%%%%%%%%%%%%%%%%%%%%%%%%%%%%%%
\begin{figure}[bt]
\begin{center}
\includegraphics[width=\columnwidth]{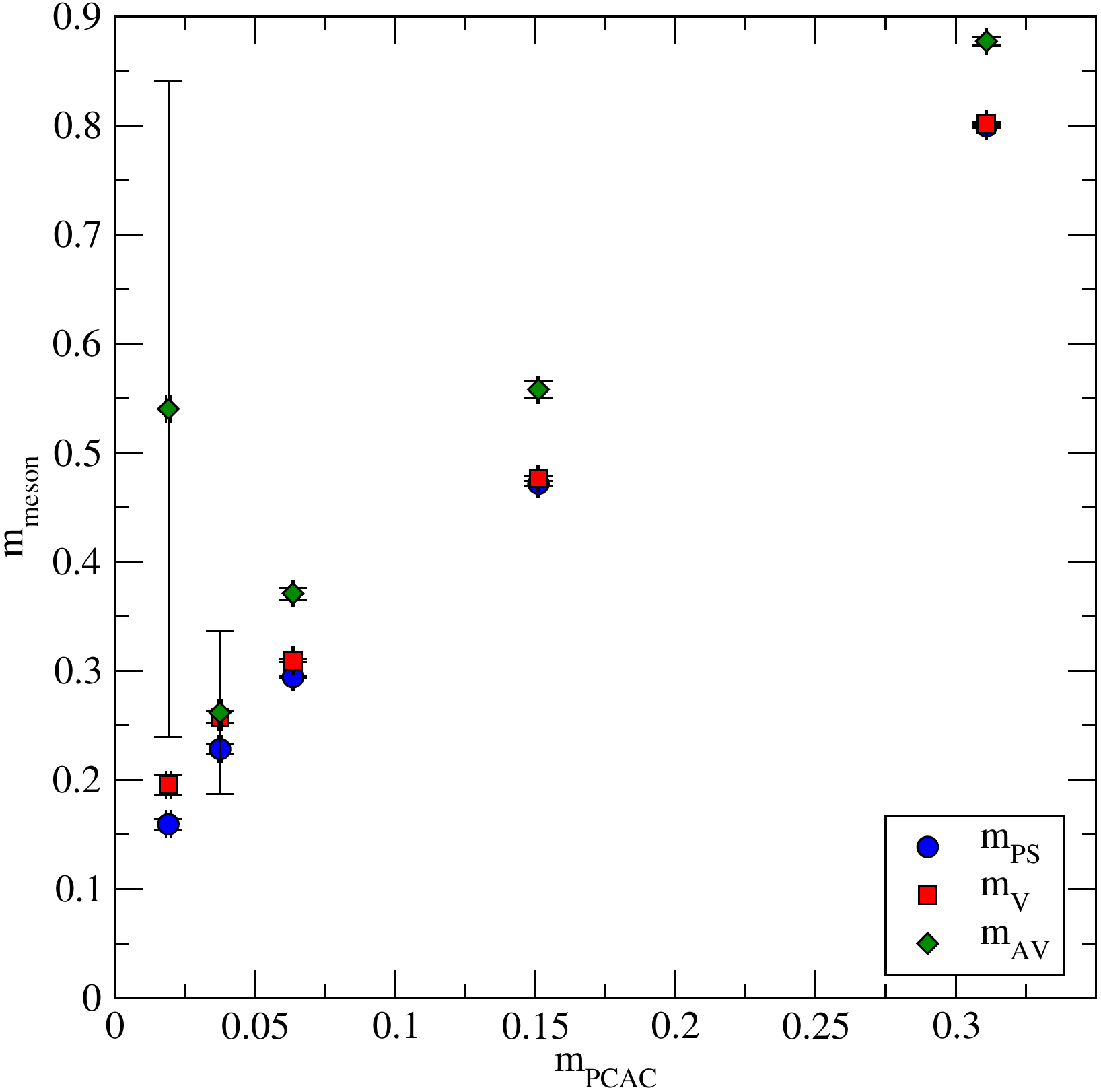}
\caption{Pseudoscalar, vector, and axial vector meson masses measured on a $24^3\times64$ lattice at $\beta=7$.}
\label{fig:meson_masses}
\end{center}
\end{figure}
%%%%%%%%%%%%%%%%%%%%%%%%%%%%%%%%%%%
Fig.~\ref{fig:meson_masses} shows the pseudoscalar, vector, and axial vector meson masses measured on a $24^3\times64$ lattice at $\beta=7$ as the bare quark mass is decreased towards the critical value. At the lightest quark mass the pseudoscalar meson has a mass of about $m_\smc{PS}\simeq 0.15$ in lattice units. This means that $m_\smc{PS}\cdot L \simeq 3.6$ which is where finite volume effects starts to become relevant. At large quark masses the vector and pseudoscalar are degenerate with the common mass increasing linearly with the quark mass. At smaller masses the vector meson becomes heavier than the pseudosclar. This is consistent with dynamical generation of a chiral scale. To see this more clearly the ratio of the vector and the pseudoscalar masses have been plotted in Fig.~\ref{fig:PS_V_ratio}. Indeed the mass ratio approaches unity for large quark masses. However, when approaching the chiral limit the ratio increases signaling chiral symmetry breaking. In fact this result is consistent with the expectation that if spontaneous symmetry breaking occurs the vector meson remains massive whereas the pseudoscalar meson is massless. A diverging ratio $m_\smc{V}/m_\smc{PS}$ therefore indicates chiral symmetry breaking. This is the trend we observe in Fig.~\ref{fig:PS_V_ratio}. However to nail this conclusion more studies have to be performed. 

The axial mass in the chiral limit is poorly determined Fig.~\ref{fig:meson_masses}. In the future we plan on improving its determination. We will then be able to use it to infer interesting properties of the chiral transition. For example one can investigate whether the axial remains (near) degenerate with the vector in the chiral regime which could signify the onset of walking dynamics \cite{Appelquist:1998xf,Appelquist:1999dq}. 

%%%%%%%%%%%%%%%%%%%%%%%%%%%%%%%%%%%
%%%%%%%%%%%%%%%%%%%%%%%%%%%%%%
\begin{figure}[bt]
\begin{center}
\includegraphics[width=\columnwidth]{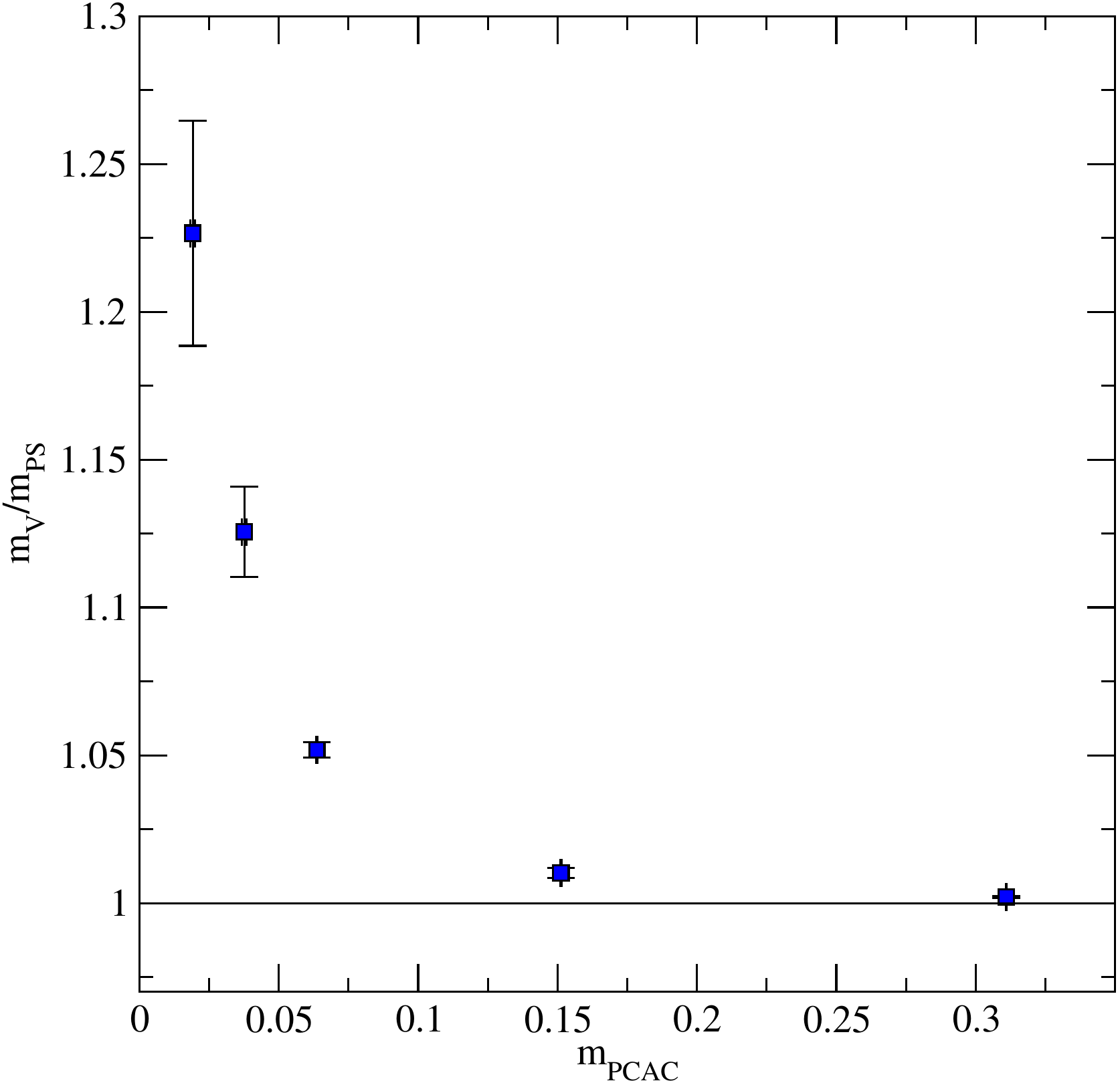}
\caption{Ratio between pseudoscalar and vector meson masses measured on a $24^3\times64$ lattice at $\beta=7$.}
\label{fig:PS_V_ratio}
\end{center}
\end{figure}
%%%%%%%%%%%%%%%%%%%%%%%%%%%%%%
\begin{figure}[bt]
\begin{center}
\includegraphics[width=\columnwidth]{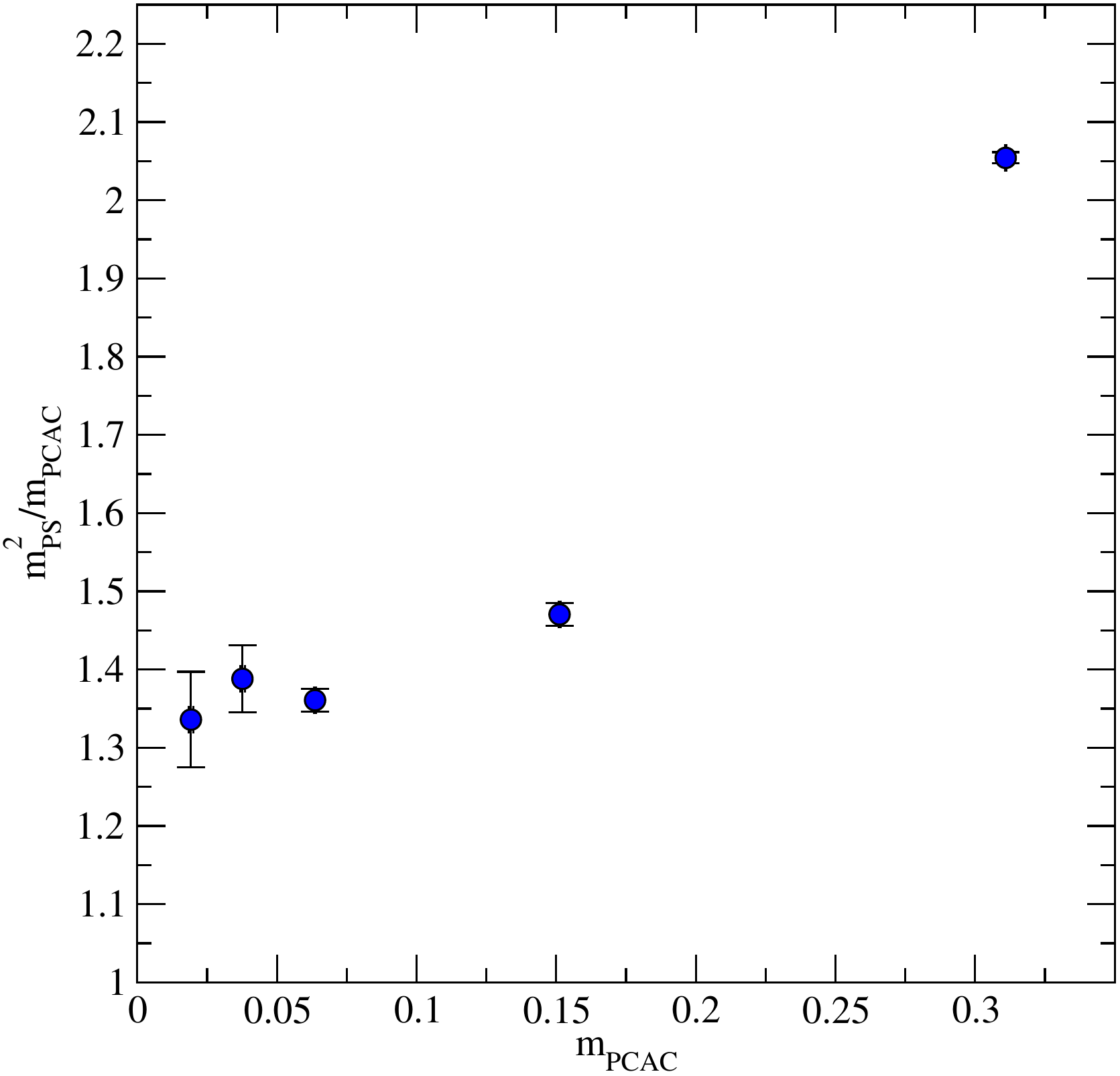}
\caption{Psudoscalar mass squared divided by the quark mass measured on a $24^3\times64$ lattice at $\beta=7$.}
\label{fig:PS2_PCAC}
\end{center}
\end{figure}
%%%%%%%%%%%%%%%%%%%%%%%%%%%%%%

%%%%%%%%%%%%%%%%%%%%%%%%%%%%%%
\begin{figure}[bt]
\begin{center}
\includegraphics[width=\columnwidth]{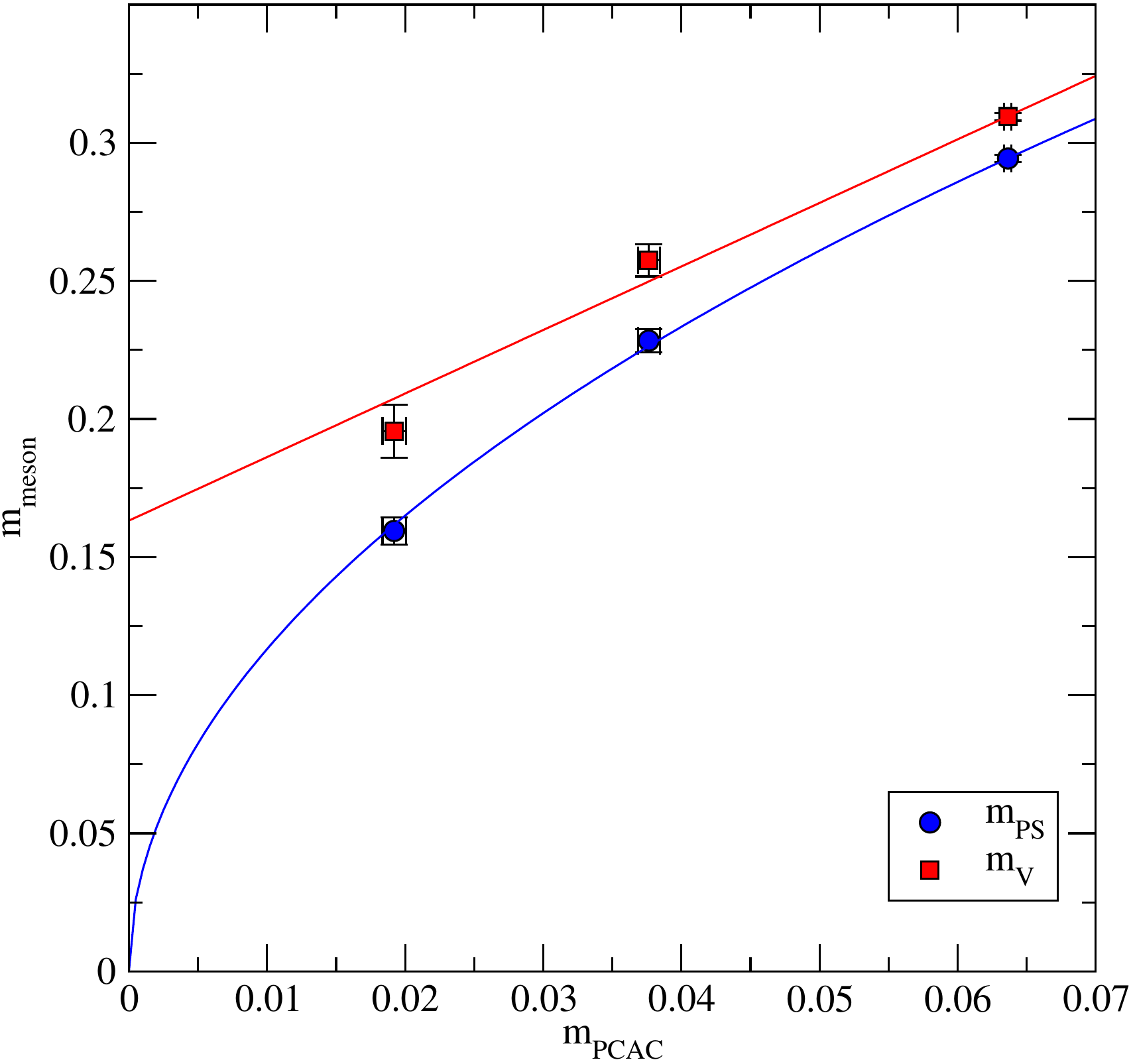}
\caption{The chiral fits to the pseudo scalar and vector meson masses on a  $24^3\times64$ lattice at $\beta=7$.}
\label{fig:meson_fits}
\end{center}
\end{figure}
%%%%%%%%%%%%%%%%%%%%%%%%%%%%%%%%%%%

\begin{table}
  \begin{tabular}{llll}
    \hline
    meson fit & fit function\;\;\; & best parameter\;\;\; & $\chi^2/$dof   \\
    \hline
    \hline
    ps chiral &   $a\sqrt{m}$ & $a=1.167(6)$ & 0.43/2 \\
    ps conformal & $am$ & $a=4.69(3)$ &  364/2 \\
     \multirow{2}{*}{ps alt. 1} &  \multirow{2}{*}{$a + bm$} & $a=0.111(6)$  & \multirow{2}{*}{6.4/1} \\
     & & $b=2.9(1)$ & \\
     \multirow{2}{*}{ps alt. 2\;\;\;} &  \multirow{2}{*}{$a + b\sqrt{m}$} & $a=-0.001(10)$  & \multirow{2}{*}{0.41/1} \\
     & & $b=1.17(4)$ & \\
     \hline
      \multirow{2}{*}{vector chiral} &  \multirow{2}{*}{$a + bm$} & $a=0.16(1)$ &  \multirow{2}{*}{3.3/1}  \\
      & & $b=2.3(2)$ & \\
      vector conformal\;\;\; & $am$ & $a=4.91(3)$& 273/2 \\
      vector alt. 1 & $a\sqrt{m}$ & $a=1.231(6)$ & 18/2 \\
      \multirow{2}{*}{vector alt. 2\;\;\;} &  \multirow{2}{*}{$a + b\sqrt{m}$} & $a=0.07(2)$  & \multirow{2}{*}{0.69/1} \\
     & & $b=0.96(7)$ & \\
     \hline
  \end{tabular}
  \caption{Different types of fit functions in the chiral regime for the data with $m$ identified with the $m_\smc{PCAC}$.} \label{tbl:fits}
\end{table}

To extract further properties of the theory we analyze in more detail the functional dependence of the pseudocalar mass on the quark mass. It is well known that, for this kind of theories, spontaneously broken chiral symmetry leads to the Gell-Mann--Oakes--Renner relation \cite{GellMann:1968rz} 
\begin{align}
m^2_\smc{PS} \simeq  \Lambda m_\smc{PCAC} \, ,
\label{GMOR}
\end{align}
valid in the chiral limit, where $\Lambda = -2\langle \bar \psi \psi\rangle / f_\smc{PS}^2$ is a dynamically generated scale. For conformal theories the behavior is different \cite{Sannino:2008nv, Sannino:2008pz}. In \cite{Sannino:2008pz} it was also shown that the instanton contributions to conformal chiral dynamics can be neglected when the anomalous dimension of the mass operator is less than one. This property has been investigated and confirmed via lattice simulations in \cite{Bennett:2012ch}. A clever separation of the ultraviolet and infrared modes presented in \cite{DelDebbio:2010ze,DelDebbio:2010jy} led to a better understanding of the conformal chiral scenario but without discussing the instanton contributions \cite{Sannino:2008pz}. Building upon these results an interesting method to determine the anomalous dimension of the fermion masses was put forward in \cite{Patella:2011jr}. 
To sum up, for a conformal scenario the dynamical scale $\Lambda$ mutates into a fermion-mass dependent quantity \cite{Sannino:2008pz} and therefore $m^2_\smc{PS}$ must vanish as $m^2_\smc{PCAC}$. In Fig.~\ref{fig:PS2_PCAC} we plot the ratio $m^2_\smc{PS}/ m_\smc{PCAC}$ for decreasing fermion mass. We see that the ratio approaches a constant for vanishing fermion masses which is consistent with the chiral symmetry breaking scenario \eqref{GMOR}.

In  table \ref{tbl:fits} we report the fit to the data for the dependence of the pseudoscalar mass as well as the vector mass as function of the $m_\smc{PCAC}$  within the believed chiral regime of the theory.   This corresponds to the three lowest values of $m_\smc{PCAC}$ where the ratio $m^2_\smc{PS}/m_\smc{PCAC}$ becomes roughly constant  as shown in Fig.~\ref{fig:PS2_PCAC}. The data points used for the chiral fits in the table  are shown in Fig.~\ref{fig:meson_fits}. The best fit curve, determined by the lowest $\chi^2/{\rm dof}$, for the pseudoscalar mass corresponds to the first line of the table which is in agreement with the GMOR expectation. It is remarkable that by even allowing for an offset of the mass value in the chiral limit the best fit demands the offset to vanish, see the last line of the table. We have tried also to test the possibility that the pseudoscalar mass vanishes linearly with the fermion mass and found that this is highly disfavored. If the theory would have been conformal we would have expected this case to fit much better. 

Similarly, by fitting the vector masses dependence on the fermion mass, in the lower part of table \ref{tbl:fits}, we observe a reasonable agreement with the expected chiral behavior of the theory.  The two best fits correspond to the first and last line of the lower part of the table. We would have expected the first line to yield a better fit if chiral symmetry breaks like in ordinary Quantum chromodynamics. We believe that for this case more statistics is needed to resolve which of the two cases is actually realized given that the data cannot yet differentiate between the two. As for the pseudoscalar case the would be conformal case is highly disfavored (see second line of the lower part of the table). 

Using the identity for the hadronic correlators  \eqref{CorrId} we can immediately infer the baryonic diquark masses. 

\section{Conclusions}
\label{conclusion}
Orthogonal lattice gauge theories with dynamical fermions have so far been {\it terra incognita}. However, as explained in the introduction, these theories can be relevant for models of dynamical electroweak symmetry breaking as well as for the construction of interesting dark matter candidates. Furthermore to have a deeper understanding of strong dynamics it is essential to gain information on different gauge theories.  We have chosen to start investigating the orthogonal gauge groups dynamics with a phenomenologically relevant example, i.e. the SO($4$) gauge theory with two Dirac flavors transforming according to the vector representation of the group. 

We have uncovered the lattice phase diagram and shown that there is a novel phase separation phenomenon at small volumes which persists even in the pure gauge case. We have shown that the phase separation can be circumvented  and the chiral regime of the theory studied using large but still feasible lattices. 

Finally we investigated the spectrum of the theory for the pseudoscalar, vectors and axial vectors. The results for the spectrum are consistent with chiral symmetry breaking and strongly disfavor a conformal behavior.

%\acknowledgments

\appendix
\section{Diquark correlators}\label{AppCorrelator}
A generic mesonic correlator will have the form
\begin{equation}
c_{\bar \psi \psi'}^{(\Gamma )}(x-y)= \tr \left( \left[  \bar \psi(x) \Gamma \psi'(x) \right]^\dagger  \bar \psi(y) \Gamma \psi'(y)  \right) \, ,
\end{equation}
and the baryonic diquark correlator will have the form
\begin{equation}
c_{ \psi \psi'}^{(\Gamma )}(x-y)=  \tr \left( \left[ \psi^T(x) C \Gamma \psi'(x) \right]^\dagger \psi^T(y) C \Gamma \psi'(y) \right)  \, .
\end{equation}
Rewriting the diquark correlator slightly gives
\begin{equation}
c_{\psi \psi'}^{(\Gamma )}(x-y)
%&=
%\tr \iint_{\vec x, \vec y}\left[  q^T(x) C \Gamma q'(x) \right]^\dagger  q^T(y) C\Gamma q'(y)  \nonumber\\
=
 \tr \left(  \Gamma \psi'(y) \bar \psi'(x) \gamma^0 \Gamma^\dagger C^\dagger(\gamma^0)^T  \left[ \psi(y)\bar \psi(x) \right]^T C \right)  \, .\label{CqqIdTemp1}
 \end{equation}
Now we can invoke two identities
\begin{align}
(\gamma^\mu)^T&=-C\gamma^\mu C^\dagger \, , \\
\psi(x)\bar \psi(y) & = C^\dagger \left[  \psi(y)\bar \psi(x)   \right]^T C \, .\label{qbqsym}
\end{align}
The latter identity follows from the symmetry of the Dirac matrix given in \eqref{DiracMatrixSymmetry} along with $\gamma^5$-hermiticity $\gamma^5(\slashed D+m) \gamma^5 = (\slashed D + m)^\dagger$. The identity \eqref{qbqsym} extend to the Wilson lattice formulation of the Dirac matrix. This is demonstrated for pseudoreal representations in the appendix of \cite{Lewis:2011zb}. 
Invoking the identities in the expression for the diquark correlator \eqref{CqqIdTemp1} we have
\begin{align}
\begin{split}
c_{\psi \psi'}^{(\Gamma )}(x-y)
&=
\tr \left(\Gamma \psi'(y) \bar \psi'(x) \gamma^0 \Gamma^\dagger \gamma^0  \psi(x)\bar \psi(y)  \right) =c_{\bar \psi \psi'}^{(\Gamma )}(x - y)
\, .
\end{split}
\end{align}
A similar derivation holds for the antiparticles leading to the identity \eqref{CorrId}.

\end{document}